\documentclass[structabstract]{aa}  
\usepackage{graphicx}
\usepackage{epstopdf}
\usepackage{subfigure}
\usepackage{longtable}
\usepackage{footmisc}
\usepackage{amsmath}
\usepackage{txfonts}
\usepackage{upgreek}
\usepackage{natbib}
\bibpunct{(}{)}{;}{a}{}{,}
\usepackage{url}
\begin{document}
   \title{Polarimetric coronagraphy of  BD$+$31$\degr$643
   \thanks{Based on observations made with the Nordic Optical Telescope, operated
on the island of La Palma jointly by Denmark, Finland, Iceland,
Norway, and Sweden, in the Spanish Observatorio del Roque de los
Muchachos of the Instituto de Astrofisica de Canarias. }
 } 

   \author{G. Olofsson
          \inst{1}
          \and
           R. Nilsson
          \inst{1}
          \and       
           H.-G. Flor\'en
          \inst{1}
          \and
          A. Djupvik
          \inst{2}
	 \and
          M. Aberasturi
          \inst{3}
	 }
          
   \offprints{G. Olofsson}

   \institute{Department of Astronomy, Stockholm University, AlbaNova University Center, Roslagstullsbacken 21, SE-106 91 Stockholm, Sweden\\
              \email{olofsson@astro.su.se, ricky@astro.su.se, floren@astro.su.se}
          \and
           Nordic Optical Telescope, Apado 474,38700 Santa Cruz de La Palma, Spain\\
           \email{amanda@not.iac.es}   
	\and
	   Department of Astrophysics, Centre for Astrobiology (CAB, CSIC-INTA), ESAC
Campus, P.O. Box 78, 28691 Villafranca de la Ca\~nada, Madrid, Spain\\
         		 }

   \date{Received \today; accepted Month Day, Year}

 
  \abstract
   {The binary B5\,V star BD$+$31$\degr$643 exhibits a disk-like structure detected at optical wavelengths. Even though the feature is well centered on the star, it has been argued, based on \emph{Spitzer } observations, that the feature is a filament not directly associated to the binary star.}
   {The purpose of the present paper is to investigate whether polarization imaging may provide evidence either for or against the disk hypothesis. In addition,  we aim at clarifying whether there might be any additional close companion to the binary star.}
   {We used the coronagraph \emph{PolCor} in its polarization mode in combination with an EMCCD camera allowing short unit exposure times. As a result of shift-and-add and frame selection, the spatial resolution is improved compared to traditional CCD imaging. In order to possibly reveal an additional stellar companion, we used high resolution spectroscopy in the optical and high spatial resolution imaging in the near-IR. }
   {The disk/filament is much better seen in polarization; it is narrow and a line drawn along the ridge passes within a second of arc from the star. The degree of polarization is high ($\approx$50\% after correction for the extended component of the reflection nebula) which means that the disk/filament must be approximately at the same distance as the star. Although we confirm that the feature is much brighter south-east than north-west of the star, the evidence that the feature is physically connected to the star is strengthened and suggests that we are witnessing the destruction process of an accretion disk. Our spectroscopy shows that at least one of the stars is a spectroscopic binary. We were, however, not able to spatially resolve  any stellar component in addition to the two well separated stars.   }
   {}

	 \titlerunning{Polarimetric coronagraphy of  BD$+$31$\degr$643: a case for an extensive circumbinary disk}

   \maketitle
%

\section{Introduction}

IC 348 is a young star cluster, located next to the Perseus molecular cloud complex, with its most luminous member -- the B5$\,$V binary BD$+$31$\degr$643 --  partially embedded to create a prominent reflection nebula visible both in the optical and near-infrared (NIR). Although the cluster did not necessarily form in the cloud, its most massive star(s) is undoubtedly interacting with it through stellar winds and UV radiation \citep[][ and references therein]{Arce2010}. \emph{Spitzer} maps reveal that the IR emission surface brightness peak of the cloud coincides with the location of BD$+$31$\degr$643 \citep{Schmeja2008,Muench2007}, and that an incomplete spherical shell (with a diameter of $\sim$200{\arcsec}) surrounds the binary \citep{Rebull2007}.

One of the most debated features in the nebulosity is the bright linear streak first imaged with optical coronagraphy at the University of Hawaii 2.2\,m telescope by \citet{Kalas1997}, centred on BD$+$31$\degr$643, with an extent of 6$\,$600$\,$AU. It was originally believed to originate from light scattered in a huge circumstellar debris disk, but subsequent imaging of the region has cast doubt on the nature of the streak. Although a narrow disk-like feature with the same position angle across the star was seen with \emph{Spitzer} in extended 24{$\,\upmu$}m emission, \citet{Rebull2007} dismissed the disk theory based on several arguments: 1) the scattered optical disk would be an order of magnitude larger than any known debris disk; 2) the size at 24$\,\upmu$m is three times bigger that the optical size; 3) no excess IR \citep[or submillimetre,][]{Enoch2006} emission is seen in the spectral energy distribution (SED) of the star \citep[although initially suggested by ISO SWS observations by][the spectrometer's large aperture was probably contaminated by extended emission]{Merin2004b}; 4) extreme 35{\degr} opening angle toward the south-east (SE); and 5) strong brightness asymmetry. The 24$\,\upmu$m surface brightness peaks 22{\arcsec} SE of the star and is four times brighter than on the corresponding north-west (NW) position. In addition, it extends to 55{\arcsec} on the SE side, but reaches 80{\arcsec} on the NW side.

\citet{Rebull2007} instead suggest that the linear structure observed in the cavity cleared by BD$+$31$\degr$643 may be caused by a filamentary interstellar cloud---a common substructure in star forming molecular clouds \citep[see, e.g.,][]{Williams2000} -- which coincidentally crosses behind or in front of the star. The brightness asymmetry would be explained by the filament's viewing angle and closer approach to the SE side of the star.  The physical properties of the star BD$+$31$\degr$643 are summarized in Table\,\ref{table:data}.


\begin{table}
\caption{Coordinates and physical properties of BD$+$31$\degr$643 (also commonly denoted HD 281159).}             
\label{table:data}      
\centering          
\begin{tabular}{ l l l}     
\hline\hline       
Parameter & Value & Ref.\\ 
\hline
   Position (J2000) & R.A.= 03h$\,$44m$\,$34.19s & 1\\
          & Dec. = +32{\degr}09{\arcmin}46.14{\arcsec} & 1 \\          
   Distance\tablefootmark{a}, $D$ & $300\,$pc & 2 \\
   Spectral type and luminosity class & B5$\,$V \& B5$\,$V binary &  \\
   Effective temperature, $T_{\textnormal{eff}}$ & $15\,400\,$K & 3 \\
   Stellar mass, $M_{*}$ & $6.0{\pm}1.0\,M_{\odot}$ & 4 \\
   Surface gravity, log$\,g$ & $3.75{\pm}0.05\,$cm$\,$s$^{-2}$ & 5 \\
   Age & 1.3$\,$Myr & 4 \\
\hline                  
\end{tabular}
\tablebib{
(1)~vanLeeuwen2007; (2)~\cite{Herbst2008}; (3)~\citet{Merin2004b}; (4)~\citet{Preibisch2001}; 
(5)~\citet{Montesinos2009}
}
\tablefoot{
\tablefoottext{a}{The distance measured from the mean parallax of IC$\,$348 cluster members is  ($\sim$260$\,$pc), while the distance  derived from the HR diagram is  greater ($\sim$320$\,$pc) \citep[see][and references therein]{Haisch2001}.}}
\end{table}

\subsection{Aims of this study}
The scattered light imaged by \citet{Kalas1997} and the extended IR emission  seen by \citet{Rebull2007} both reveal a disk-like feature crossing the binary star BD$+$31$\degr$643, but results  remain inconclusive on whether or not it actually is a disk. Multicolor \emph{aperture} polarimetry has shown a polarization that is parallel to the major axis of the projected disk feature and perpendicular on the sky to the projected direction of the local magnetic field \citep{Andersson1997}. Our goal was to make the first polarimetric \emph{mapping} of the structure, with the highest spatial resolution and contrast to date, using polarimetric coronagraphy with PolCor at the 2.6-m Nordic Optical Telescope (NOT), complemented with high-resolution optical and IR spectra, and high spatial resolution IR imaging, in order to discriminate between its filamentary or circumstellar disk nature, and possibly resolve additional stellar components.


\section{Observations and data reduction}\label{sec:obs}

\subsection{Imaging with PolCor at the NOT}

PolCor is a stellar coronagraph with a polarization mode. It is equipped with a 512$\times$512 EMCCD camera (Andor IXON) which allows for very high frame rates (typically 10--30 Hz, but occasionally much faster on subframes) without loss in sensitivity. It also makes it possible to use a short observation cycle (normally 20\,s) taking images through a polarizer at the four positions (0, 45, 90 and 135\degr). We also include a dark measurement in the unit cycle to keep track on possible drifts. All frames are stored on disk, and the post-processing includes frame selection and aligning of the frames (shift-and-add). There is a choice of three different sizes for the occulting disks (1.5, 3 and 6\arcsec  diameter) each with three choices of damping (ND 2, 3.5 and 5). The optical density of the occulting disk is chosen to avoid saturation and still make the star visible in each frame, which is important for accurate centering of the star as well as for the post-processing. In order to cancel the diffraction from the support blades of the secondary mirror, a computer controlled, rotating Lyot stop is used, which shades the image of the secondary mirror as well as these blades. For further technical details, see \cite{Ramstedt2011}.

The observations presented in the present paper were carried out on Oct 8 (V filter) and Oct 10 (R and H$\alpha$ filters) 2007 and  BD$+$31$\degr$643 was observed in the V and R filters, each in 100 cycles (15000 unit frames) representing on-target observation time of 5 min for each filter and polarizer position. The seeing conditions were variable (0.8 - 1.1$\arcsec$). The sensitivity in surface brightness, normally limited by the shot noise from the sky background or (close to the star) the stellar point-spread-function (PSF), is in this case limited by the shot noise as well as small scale structures of the reflection nebula.

The PSF was determined from a standard star and polarization standards were used for the polarimetric calibrations.

\subsection{Reducing PolCor data}

The reduction of the PolCor data briefly includes the following steps:
\begin{enumerate}
\item  subtract (for each cycle) the average dark  frame from each individual frame
\item  calculate the position and the sharpness of the star for each frame
\item  exclude frames with poor sharpness
\item  co-add the sharp frames for each polarizer position by a shift-and-add procedure
\item  fine-tune the centering of co-added images for the four polarizer positions to exactly coincide
\item  subtract the PSF from each of the four images (only needed if the star is polarized, which is the case for BD$+$31$\degr$643
\item  calculate the Stokes $Q$ = image\,0$\degr$--image\,90$\degr$ and Stokes $U$ = image\,45$\degr$--image\,135$\degr$
\item  calculate the Stokes $I$ =  ( image\,0$\degr$ + image\,90$\degr$ + image\,45$\degr$ + image\,135$\degr$)/2.
\item  calculate the \emph{polarized power, $P_{\mathrm{pwr}}$ } =$\sqrt{Q^2 + U^2}$
\item  calculate the \emph{polarization degree,  p} = 100\,$P_{\mathrm{pwr}}/I $  (in \%)
\item  calculate the \emph{angle of polarization}, $\theta$ = 0.5\,$\arctan(U/Q)$
\end{enumerate}

\subsection{Spectroscopy with FIES at the NOT}

FIES is a fibre-fed high resolution ($R = 67000$) spectrograph with a fibre diameter corresponding to 1.3$\arcsec$ The observations were carried out Aug 30, 2010. The software package FIEStool version 1.3.2 was used for the reductions.\footnote{For details, see 
\url{http://www.not.iac.es/instruments/fies/fiestool/FIEStool.html}.} 

\subsection{Intermediate and low resolution spectroscopy at the INT}

The low resolution spectroscopy was taken with the Isaac Newton
Telescope (INT) (La Palma, Spain) equipped with the Intermediate
Dispersion Spectrograph (IDS). Two sets of observations have been
analyzed in this paper: the first one corresponds to those obtained by
the EXPORT consortium in 24--28 Oct 1998, the wavelength coverage was
5800--6700 \AA, with a resolving power of $\sim\!6000$ \citep[see][for further details]{Mora2001}. The second set was obtained on 13 Jan
2012, with a similar resolution but slightly different configuration,
covering the interval 3800--8000 \AA; in this case the position angle
of the slit allowed us to include both components of the spectroscopic binary (see
Fig. 1), but unfortunately the seeing was not good enough to
separate them.

\subsection{Near-IR imaging and spectroscopy with NOTCam at the NOT}

NOTCam is a combined camera/spectrometer for the near-IR region.\footnote{For a detailed description, see \url{http://www.not.iac.es/instruments/notcam/}} The instrument was used in its high resolution mode, both for the imaging and the spectroscopy. For imaging this means a pixel scale of 0.078$\arcsec$/pixel and for the spectroscopy a resolving power of $R = 5500$. The imaging was carried out in a standard way with 3$\times$3 grid points observed in 3 cycles using the K$_s$ and the H$_2$ S1 filters. The spectroscopy was also carried out in a standard fashion, moving the telescope back and forth in the slit direction (ABBA). The slit was oriented along the position axis of the binary (16 degrees). The observations were carried out on Sept 30, 2010 under good seeing conditions (0.6$\arcsec$ in the K band).


\section{Results}\label{sec:results}

\subsection{Imaging}

\begin{figure}
   \centering
   \includegraphics[width=9cm]{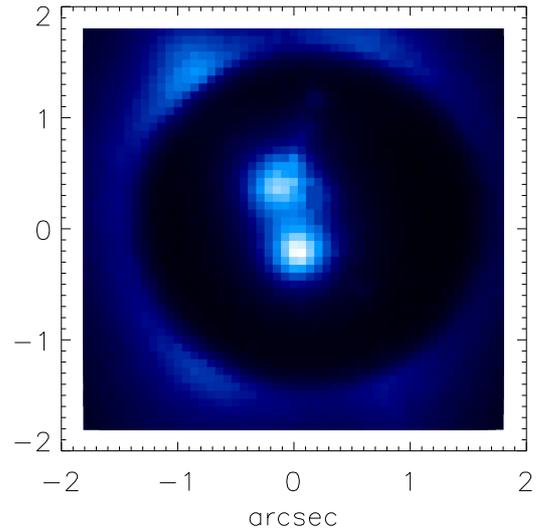}
      \caption{The binary seen through the coronagraphic mask. The separation is 0.6$\arcsec$ and as a result of the shift-and-add technique the stars are well resolved in spite of the relatively poor seeing conditions. North is up and East is left  (we use the same off-set convention in all the images in the present paper).
              }
         \label{PolCor_resolved_stars}
   \end{figure}

\begin{figure}
   \centering
   \includegraphics[width=9cm]{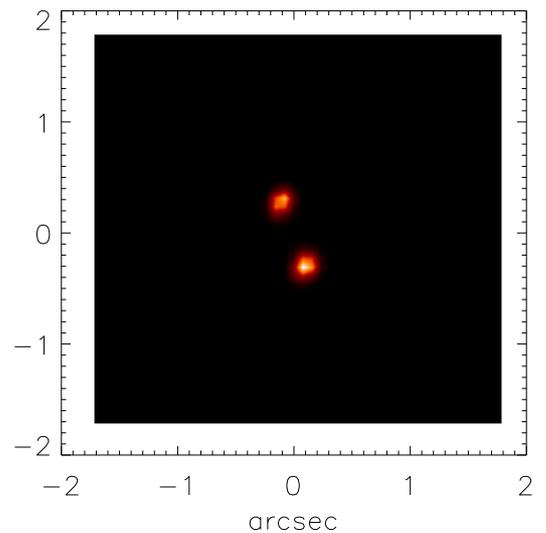}
      \caption{The binary observed with NOTCam in the K$_s$ filter and further sharpened using a maximum entropy method. No additional component is detected.
              }
         \label{Ks_image}
   \end{figure}

Apart from the polarization imaging (to be described below) we confirmed the efficiency of the shift-and-add technique, see Fig.\,\ref{PolCor_resolved_stars}. We find that the A (southern) component is 0.11$\pm$0.03 magnitudes brighter than the B component in the V band. The two stars were also well resolved in some of the individual frames taken in the K$_s$ filter. By applying a maximum entropy algorithm it is possible to further improve the sharpness and possibly resolve a third component. The result is shown in
Fig.\,\ref{Ks_image} and we find no third, spatially resolved component. The A component is 0.14$\pm$0.03 magnitudes brighter than the B component in the K$_s$ band. 

\begin{figure}
   \centering
   \includegraphics[width=9cm]{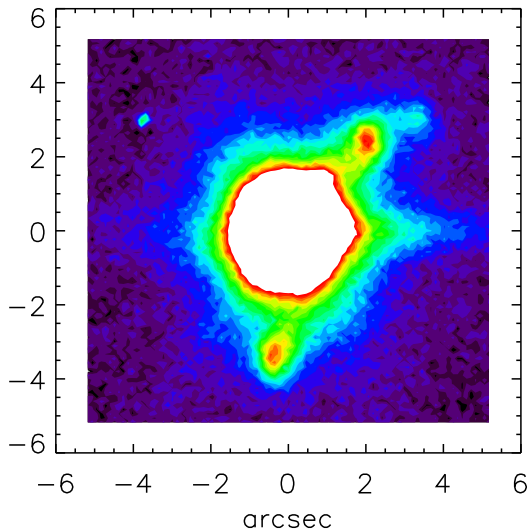}
      \caption{The A2 star (HIP\;17468) NE of  BD$+$31$\degr$643  has two point-like sources seen in this H$\alpha$ image. These features are not seen in our V or R images. The feature going straight to the west of the star is an artifact due to the readout.
              }
         \label{Halpha}
   \end{figure}

Our H$\alpha$ image shows no trace of an HII region. This is not very surprising as the stars are not quite hot enough for ionizing the surroundings \citep{Morales2001}. The only noticeable features detected in the H$\alpha$ frame are structures close to the bright star NE of BD$+$31$\degr$643 (HIP\,17468). This A2 star belongs to the young cluster population \citep{Luhman2003} and in Fig.\,\ref{Halpha} two bright knots are seen close to the star. It should be noted that this star was in the past regarded as a third component to the BD$+$31$\degr$643 system. 

Our image in the H$_2$ S1 line did not show any trace of UV fluorescence. If the disk-like feature would contain a gas component, normal for the ISM (interstellar matter), one would expect to see UV-exited H$_2$ emission, provided the hydrogen is not all in atomic form. However, as we were looking for fine spatial details, the pixel scale was not optimized for faint surface brightness. A deeper search for H$_2$ emission would be justified as it could give useful information on the dust/gas ratio of the feature.

\subsection{Spectroscopy}

\begin{figure}
   \centering
   \includegraphics[width=9cm]{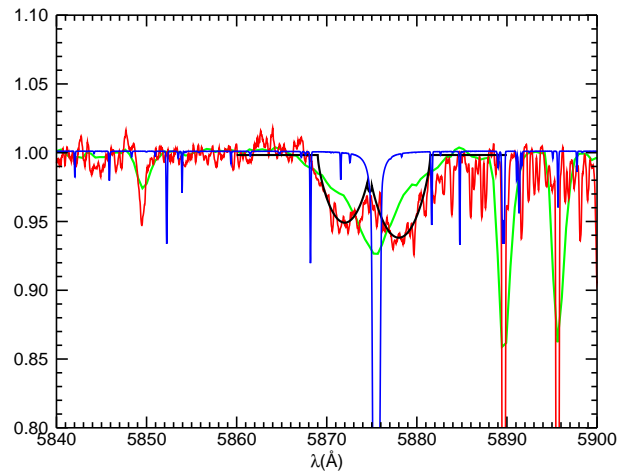}
      \caption{The high resolution FIES spectrum (red) of the region around the \ion{He}{i} 5875.62\,$\AA$  line is compared to the intermediate resolution spectrum (green) from  12 years before. The \ion{He}{i} line shape has changed dramatically. The blue spectrum represents a synthetic spectrum from an ATLAS model ($T_{\mathrm{eff}}$ = 15000\,K). The two U-shaped components represent two velocity shifted, rotating B5 stars. }
         \label{velocity}
   \end{figure}

Our high resolution spectra of  BD$+$31$\degr$643 show clear evidence of a spectroscopic binary (SB), with line profiles consisting of two U-shaped, well separated components. Looking back on observations in the past, we found published spectra, taken 11-12 years earlier with distinctly different line profiles, see Fig.\,\ref{velocity}.  To interpret the line profile in terms of two rotating stars we must first make an assumption about the radial velocity of the SB. \cite{Dahm2008} get $v_{\mathrm{rad}}$ = 13.5\,km/s as a median of cluster members and as this is close to the well defined radial velocity of the interstellar absorption lines \citep{Snow1994} we simply use the adjacent \ion{Na}{i} absorption lines as reference. In fitting the two U-shaped profiles we use the limb darkening parameter $\epsilon$ = 0.6 \citep{Royer2005} and get the following result:

SB1:   $v_{\mathrm{rad}}$ = +123$\pm$20 km/s and $v\,\sin(i) = 179\pm20$ km/s

SB2:   $v_{\mathrm{rad}}$ = -184$\pm$20 km/s and $v\,\sin(i) = 153\pm20$ km/s \smallskip

This means, assuming that the centre of gravity for the SB has the same radial velocity as the \ion{Na}{i} lines, that the mass ratio is

$m$(SB1)/$m$(SB2) = 184/123 = 1.5$\pm$0.4 \smallskip

\begin{figure}
   \centering
   \includegraphics[width=9cm]{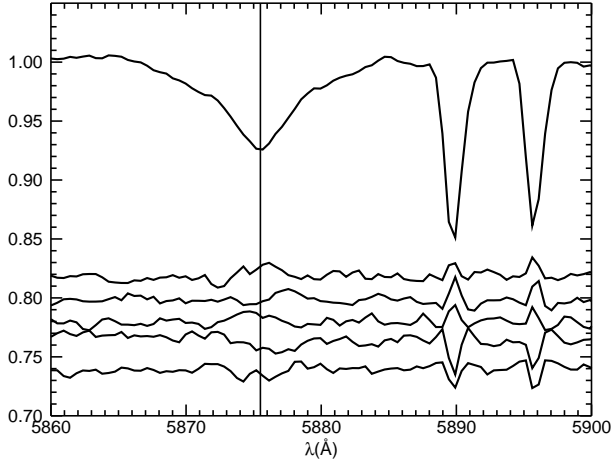}
      \caption{The \ion{He}{i}\,$\uplambda$5875.62 line observed during 5 consecutive nights in October 1998. The upper curve is the average spectrum and the lower curves display the ratios between the individual spectra and the average (with progressive offsets for visibility). The unresolved interstellar sodium lines are sensitive to the seeing and vary for that reason. The helium line is resolved and show significant time variations due to orbital motion of the SB.}
         \label{Vorbital}
   \end{figure}

So far we have ignored the fact that our FIES spectrum includes both of the visual binary stars A and B (the fibre diameter is 1.3$\arcsec$ and the separation is 0.6$\arcsec$). We first conclude that the orbital motion of A and B is small. As the projected distance is $\sim$200 AU the differential orbital motion would be at most a few km/s. Therefore at least one of the visual components is a SB. We also note that the large differential radial velocity, 307 km/s, indicates that the period of the SB must be short (probably just a few days, depending on the details of the orbital parameters). Actually, as we have access to intermediate resolution spectra during 5 consecutive nights we can test this hypothesis. In Fig.\,\ref{Vorbital} we show the average spectrum for the spectra taken on Oct 24--28, 1998 and the ratios between the individual spectra and the average spectrum. Obviously the interstellar \ion{Na}{i} doublet shows large variations, which is due to the fact that the line is unresolved and its depth is therefore seeing-dependent for the slit spectrograph. The \ion{He}{i}\,$\uplambda$5875.62 line, being broad and resolved, should not be very sensitive to the seeing and the variations should therefore be real. There are, indeed, indications of night-to-night variations, but in order to trace the velocity components, the sampling should be at least twice a night. It is also important to spatially resolve the A and B components.

As we could fit the two shifted rotation broadened \ion{He}{i} lines to the observed line ignoring the contribution of the other star, one may wonder how this contribution may influence the result. The fact that the A and B components only slightly differ in brightness and that, in addition, they have the same $V-K_s$ color index indicate that their spectral classes cannot be very different. 

\begin{figure}
   \centering
   \includegraphics[width=9cm]{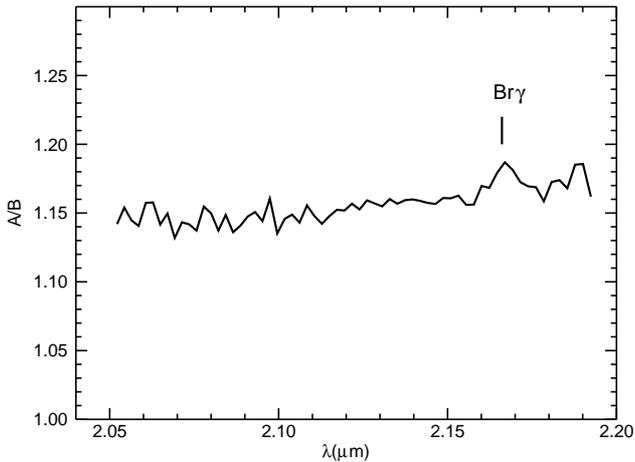}
      \caption{The ratio of the near-IR spectra for the visual binary. The slightly fainter B (north) component has a stronger Br$\gamma$ line, indicating a later spectral class.}
         \label{Brgamma}
   \end{figure}

As our near-IR slit spectroscopy spatially resolve the A and B components, we can investigate them separately. Within the uncertainty (mainly due to the cancellation of the telluric lines) the two spectra look the same with the Br$\gamma$ being the only visible feature. However, by dividing the two spectra the residual telluric features disappear and by further smoothing the spectrum to gain S/N we find that the B component has a slightly (a few per cent) stronger Br$\gamma$ line, see Fig.\,\ref{Brgamma}. This indicates that the B component has a somewhat later spectral class as one would expect as the A component is slightly brighter. 

Our recent low resolution optical slit spectra are, due to the seeing and the coarse spatial sampling, not spatially resolved. By comparing a spectral region that both includes a \ion{H}{i} and a \ion{He}{i} line, we find that the high and low resolution spectra agree quite well, except that the \ion{He}{i}\,$\uplambda$4922 line has a slightly different profile, indicating that the phase of the SB differed for the two observations. In order to gain information about the individual A and B spectra we analyze separately the spatial wings of the low resolution slit spectrum and it turns out that the ratio between the two show a tendency similar to the near-IR spectra: the \ion{H}{i} lines are stronger in the B component. We also note that the \ion{He}{i}\,$\uplambda$4472 line is (within the uncertainties) the same for the two stars. Taken together, we find that the B component is slightly later than the A component, but not by more than two subclasses in view of the identical color indices $V-K_s$.

%

\subsection{Polarization mapping}

\begin{figure}
   \centering
   \includegraphics[width=9cm]{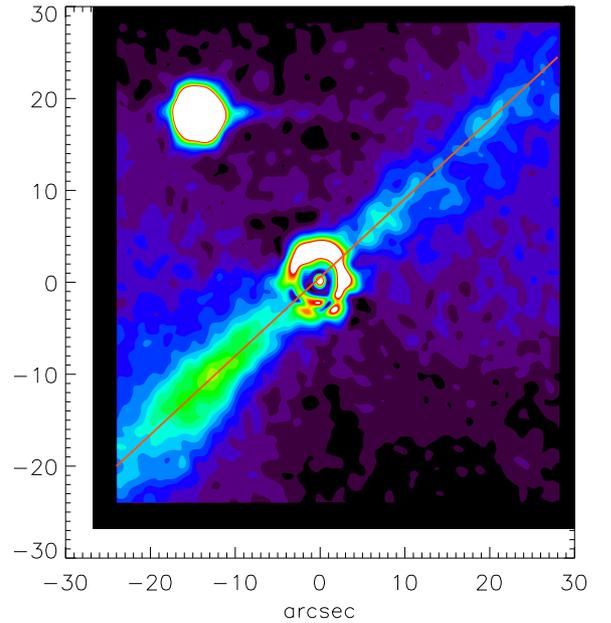}
      \caption{The polarized power in the V band. The elongated feature is quite straight and a line drawn along the ridge passes close to the central binary. The feature is however quite asymmetric, the SE part being much brighter.     }
         \label{Vpwr}
   \end{figure}

\begin{figure}
   \centering
   \includegraphics[width=9cm]{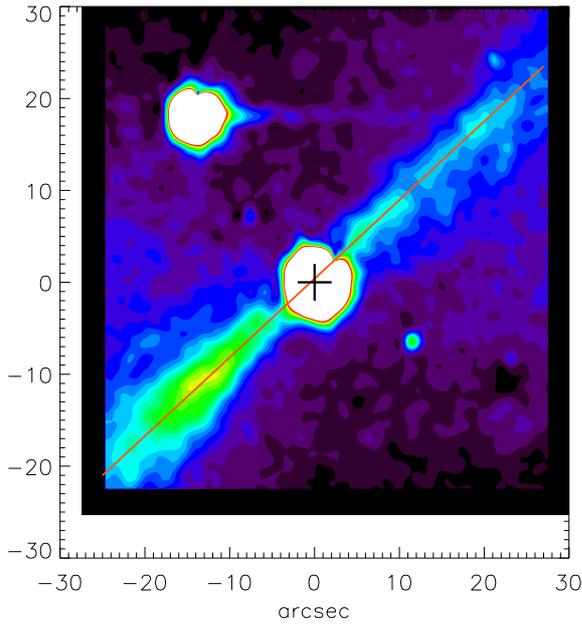}
      \caption{The polarized power observed in the R band closely mimics that seen in the V band except that the asymmetry is slightly less pronounced.   }
         \label{Rpwr}
   \end{figure}

\begin{figure}
   \centering
   \includegraphics[width=9cm]{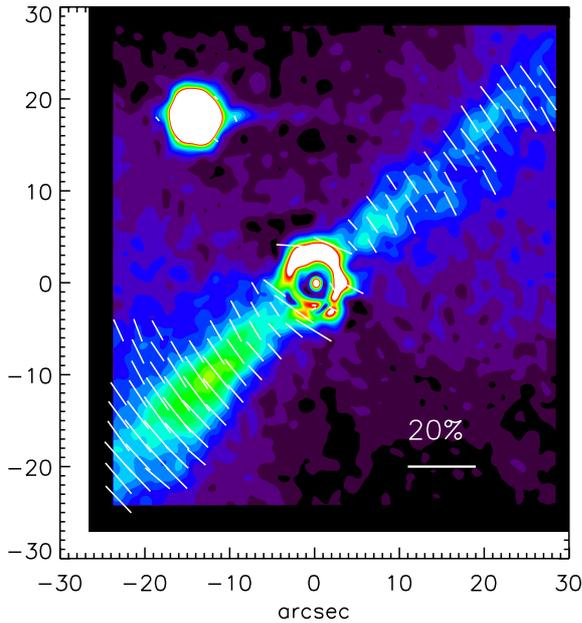}
      \caption{The polarization vectors overlaid over the polarized power image in the V band. The classical centro-symmetric pattern confirms the scattering from the central binary stars.    }
         \label{Pvectors}
   \end{figure}

\begin{figure}
   \centering
   \includegraphics[width=9cm]{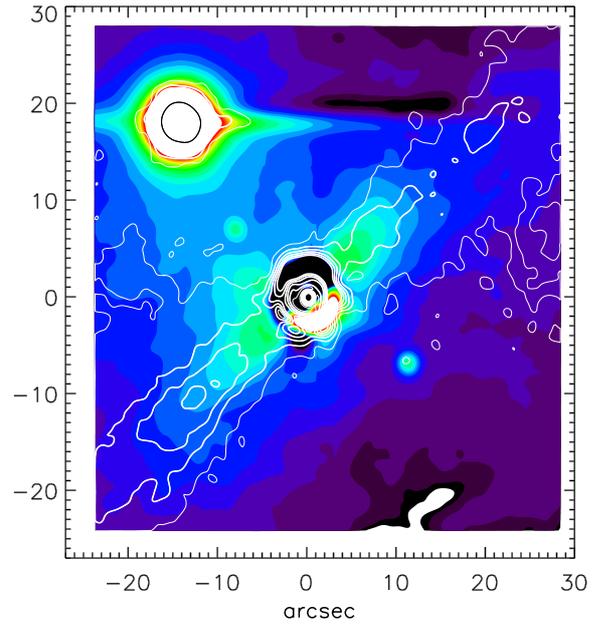}
      \caption{The $P_\mathrm{pwr}$ contours overlaid on the Stokes $I$ image in the V band. The disk-like feature is clearly seen in unpolarized light but it cannot be traced as far out as the polarized power because of the bright background nebula.    }
         \label{IplusPpwr}
   \end{figure}

Both in the V and R bands the elongated feature is well defined in the polarized power, $P_{\mathrm{pwr}}=\sqrt{Q^{2}+U^{2}} $. A straight line drawn along the ridge of the feature passes within an arcsecond from the central star, see Fig.\,\ref{Vpwr} and Fig.\,\ref{Rpwr}. As expected, the polarization vectors are perpendicular to the radius vector from the illuminating binary stars, see Fig.\,\ref{Pvectors}. The polarization degree seems to be relatively low, 10\% or less, but we must keep in mind that there is a lot of nebulosity in the region and if this is mostly in the background, the scattering angle would be far from optimal and thereby the total polarization from the background plus feature could be much lower than from the feature alone. As shown by \cite{Kalas1997}, the disk-like feature can be separated from the background nebulosity. We confirm that the feature can be identified also in unpolarized light (in our case the Stokes $I$ image) and in Fig.\,\ref{IplusPpwr} we show the $P_\mathrm{pwr}$ image contour overlaid on the Stokes $I$ image. The Stokes $I$ image shows, in addition to the disk-like feature, a blob of strong brightness to the east of the coronagraphic mask and this feature is also seen on the image taken by \cite{Kalas1997}. As this blob is less polarized, it must either be located behind or in front of the stars. 

\begin{figure}
   \centering
   \includegraphics[width=9cm]{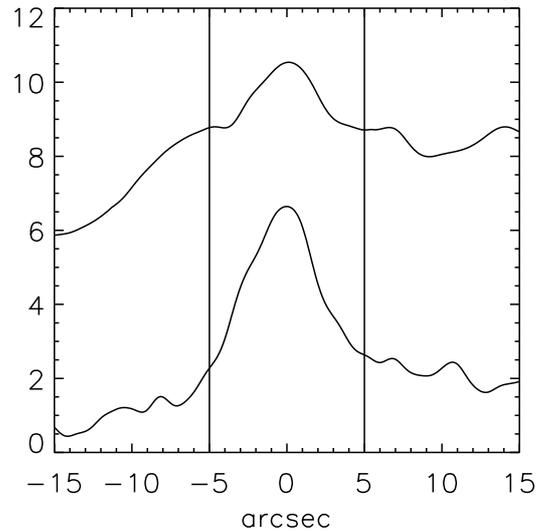}
      \caption{A cut perpendicular to the disk-like feature as observed in the V band. The upper curve represents the Stokes $I$ and the lower curve $5\times P_\mathrm{pwr}$. One unit on the y scale corresponds to 22.8 magnitudes per square arcsecond. The two vertical lines mark the positions where we define the brightness of the background nebula.   }
         \label{cut}
   \end{figure}

\begin{figure}
   \centering
   \includegraphics[width=9cm]{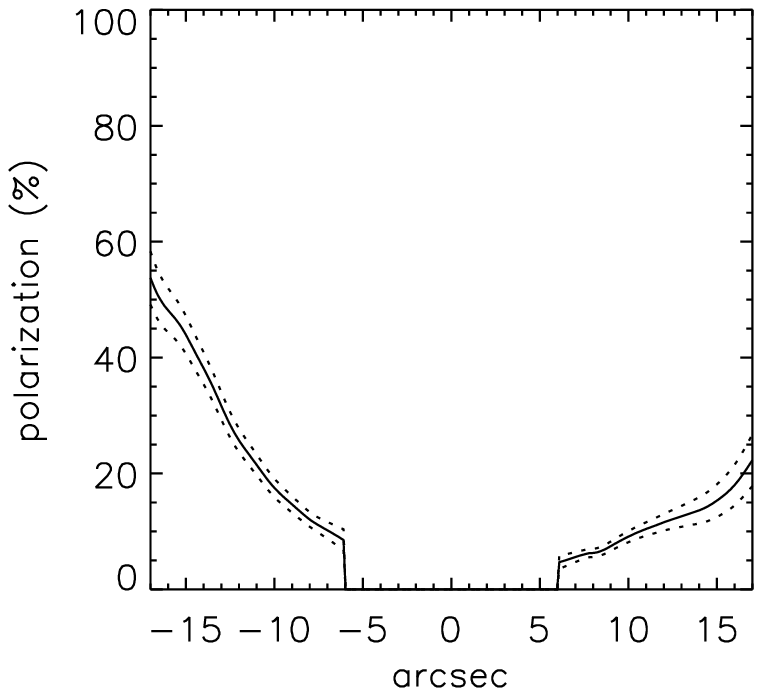}
      \caption{The degree of polarization along the disk-like feature (after correcting for the background nebula). Left is the SE offset from the star and right is the NW offset. The dotted lines represent the accuracy of the measurement.  }
         \label{diskpol}
   \end{figure}

We now, following \cite{Kalas1997}, try to subtract the background nebula and we first take a closer look at the peak polarized emission 16$\arcsec$ SE of the stars and in Fig.\ref{cut} we plot a cut through the V images in Stokes $I$ and $P_{\mathrm{pwr}}$. The narrow feature seen in the polarized power can be identified in the Stokes $I$ cut as a similar profile on top of the nebular background. We now subtract the background $\pm$5$\arcsec$ from the center of the disk-like feature (for the $P_\mathrm{pwr}$ we have to do this subtraction in Stokes $Q$ and $U$ because of the quadratic nature of $P_\mathrm{pwr}$). The result is shown in Fig.\,\ref{diskpol}. Even though there are uncertainties involved in our assumption that the background nebula can be represented by an interpolation as described above, there is a  clear indication that the disk-like feature has a very high polarization degree, in particular in the SE part. As a comparison, the disk around AU\,Mic reaches 40\% in the outer parts \citep{Graham2007}.

\section{Interpretation}\label{sec:interpretation}

\begin{figure}
   \centering
   \includegraphics[width=9cm]{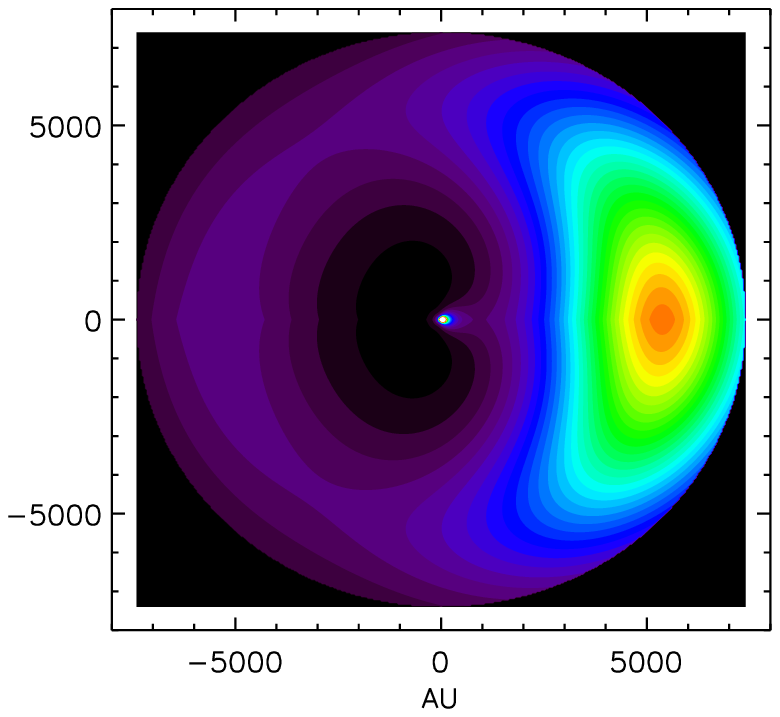}
      \caption{A circumstellar ring around the star as seen far from the right. As the phase function of the dust scattering is strongly forward-peaked, the viewer will observe a light distribution dominated by scattering from the front side of the ring.  }
         \label{donut_I}
   \end{figure}

\begin{figure}
   \centering
   \includegraphics[width=9cm]{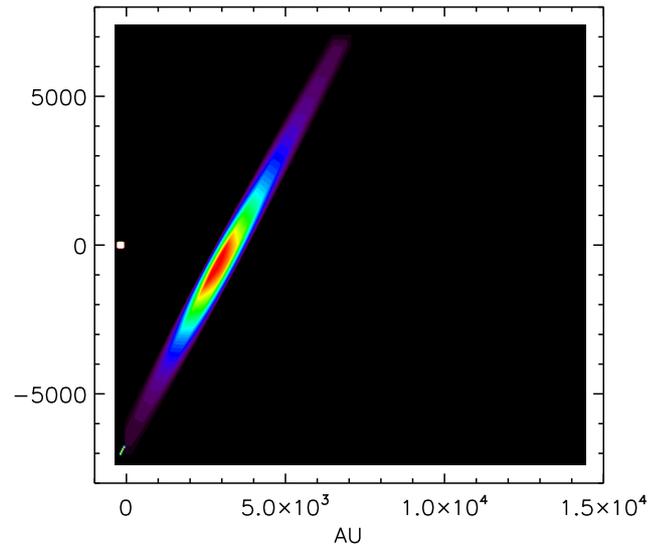}
      \caption{A filament at an angle of 63$\degr$ to the line of sight. Also in this case, the viewer (located far to the right) will observe a light distribution dominated by scattering from the front side of the filament (close to the direction of the star).  }
         \label{filament_I}
   \end{figure}

\begin{figure}
   \centering
   \includegraphics[width=9cm]{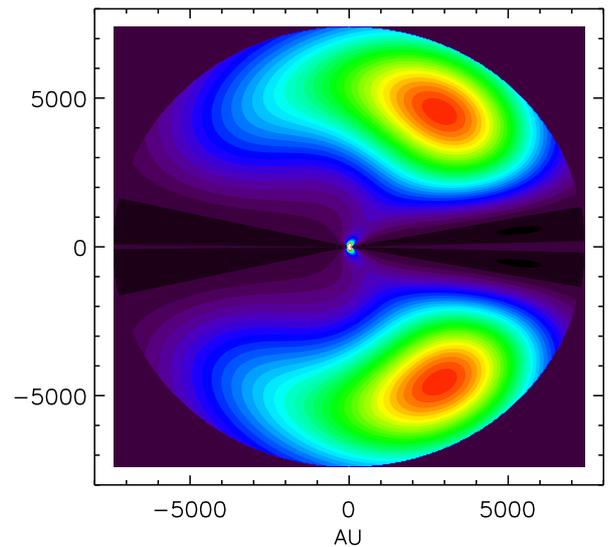}
      \caption{A circumstellar ring around the star as seen in polarized light far from the right. As the polarization strongly peaks for a scattering angle around 90$\degr$, the viewer will observe a light distribution dominated by two regions on each side of the ring.  }
         \label{donut_P}
   \end{figure}

\begin{figure}
   \centering
   \includegraphics[width=9cm]{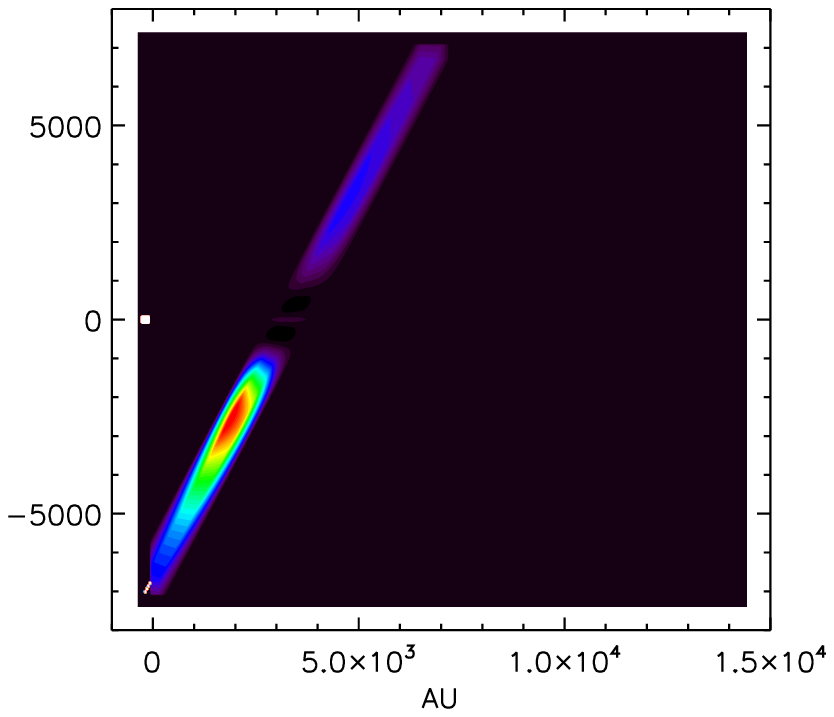}
      \caption{Two regions dominate in polarized light and the lower one (corresponding to the SE part of the disk-like feature) is much brighter -- which was the intention by tilting the filament to reproduce the observed asymmetry in polarized power.  }
         \label{filament_P}
   \end{figure}

The high degree of polarization implies that the disk-like feature cannot be located far behind or far in front of BD$+$31$\degr$643 because a high degree of polarization requires a scattering angle close to 90$\degr$. \citet{Zubko2000} have calculated the polarization properties of a MRN \citep{Mathis1977} dust particle size distribution that fits the extinction curve for the diffuse ISM. It is not clear how well this size distribution may represent the dust in the disk-like feature, in particular if the dust is the result of debris events. However, as pointed out by \cite{Kalas1997}, the scattered light is blue (as is the case for AU\,Mic but not $\upbeta$\,Pic) and this means that the scattering is dominated by small particles. Based on the calculations by \citet{Zubko2000} we investigate two different cases, a circumstellar ring and a filament located in front of and to the SE side of the the illuminating binary. Because the dust is strongly forward-scattering, the front side contributes most of the unpolarized light and the brightness profile increases strongly towards the star. This is also what is observed. The two cases are shown in Fig.\,\ref{donut_I} and \ref{filament_I}. For the polarized radiation the situation is very different and most of it is coming from the sides SE and NW of the star. For the case of a ring, see Fig.\ref{donut_P}, and the case of a filament, see Fig.\,\ref{filament_P}.

\begin{figure}
   \centering
   \includegraphics[width=9cm]{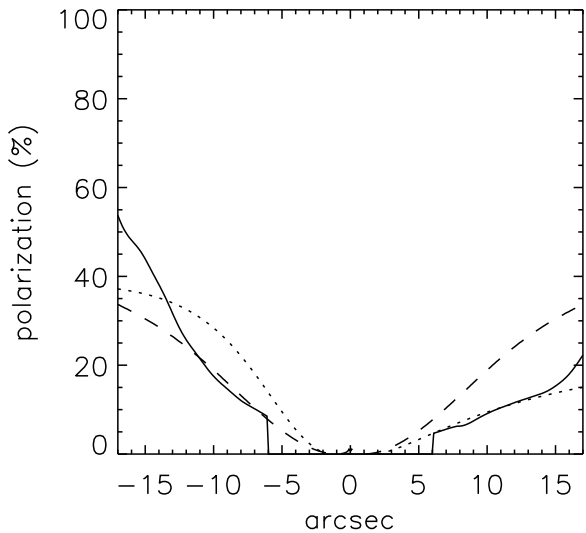}
      \caption{The observed polarization degree along the disk-like feature (full drawn line) compared to the circumstellar ring model (dashed line) and the filament model (dotted line). }
         \label{polmodels}
   \end{figure}

We are now in the position to compare the polarization profile for the two models with that deduced from the observations, see Fig.\,\ref{polmodels}. The ring case does not reproduce the asymmetry and for this reason the simple, uniform ring must be modified by lowering the density in the NW region perpendicular to the line of sight. However, the ring density has to be uniform in front of the stars to allow the projected unpolarized light distribution to be symmetric. For the case of a foreground filament our polarization model agrees reasonably well with the observations. Also in this case, a fine-tuning by allowing for density variation along the filament would be possible in order to more precisely match the observations. It seems, therefore, that both geometries (with some modifications regarding the mass distribution) are feasible. 
However, keeping in mind that the "observed" polarization profile is based on an uncertain subtraction of the background nebula, there is no point to further elaborate the models to exactly match the observations.

\section{Discussion}\label{sec:dis}

Mainly for four reasons \citep{Rebull2007} strongly question that the disk-like feature is physically connected as a circumstellar disk around  BD$+$31$\degr$643:

\begin{enumerate}
\item  The feature has huge dimensions, by far larger than any known debris disk.
\item  Spitzer MIPS mapping reveal a filament that coincides with the feature and extends even further.
\item  This filament is very asymmetric, being much brighter in the SE than in NW.
\item  There is no indication of mid- or far-IR excess at the position of the binary stars.
\end{enumerate}

\begin{figure}
   \centering
   \includegraphics[width=9cm]{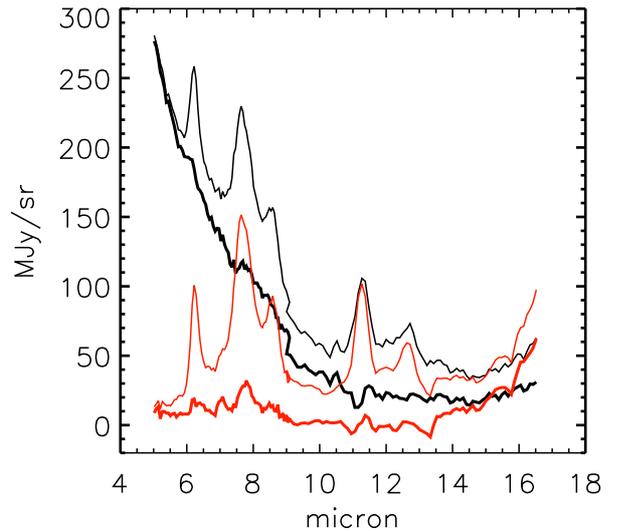}
      \caption{The ISO/CVF spectra at the $P_\mathrm{pwr}$ peak position of the disk-like feature (red, lower curve) compared to the position of the binary (black, upper curve). We also show the background-subtracted spectra of the two positions.}
         \label{CVF_Spectra}
   \end{figure}

\begin{figure}
   \centering
   \includegraphics[width=9cm]{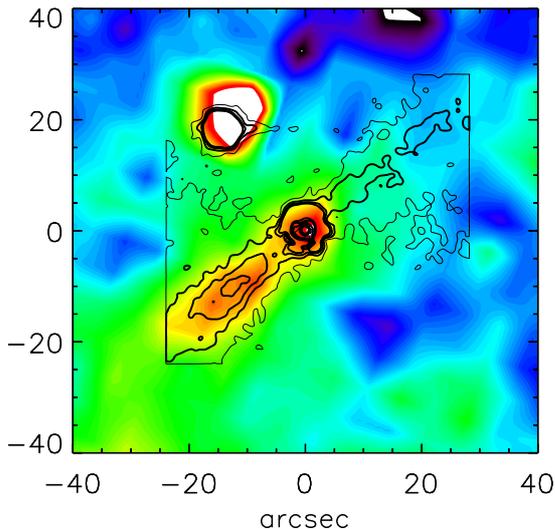}
      \caption{The $P_\mathrm{pwr}$ image in the V band overlaid as contours over the ISO/CVF image at 16\,$\mu$m. There is a very close correlation in the SE part of the feature.}
         \label{CVF_Image}
   \end{figure}

The reflection nebula around  BD$+$31$\degr$643 exhibits very bright PAH emission in the mid-IR and in order to investigate whether the disk-like feature may somehow differ from the general nebulosity, we have investigated the ISO/CVF data cube of the region. In Fig.\,\ref{CVF_Spectra} we show the CVF spectrum for the peak position of the polarized power. It has a steep rise close to the cut-off wavelength and in this respect the disk-like feature differs from the widespread PAH emission. Actually, the CVF image at 16\,$\mu$m very closely coincides with the $P_\mathrm{pwr}$ image of the disk-like feature, see Fig.\,\ref{CVF_Image}. We also note that the feature is not seen in any of the PAH peak wavelengths, so it indicates that the PAH molecules have been blown away from the vicinity of the binary stars. On the other hand, the increase in emission around 16\,$\mu$m is probably due to transient heating of PAH clusters. There is a plateau of emission often observed in the region 15--21\,$\mu$m, occasionally with distinct emission peaks \citep{Peeters2006} and for some reason the carriers of this emission have survived (or withstood the radiation pressure) for longer in the vicinity of the bright stars. The origin of the 24\,$\mu$m emission reported by \cite{Rebull2007}, which has a similar spatial feature as that at 16\,$\mu$m, is not quite clear but it is probably due to transient heated particles as the emission is wide-spread (although it peaks at the filament).

As mentioned, the disk-like feature is not seen above the background at shorter wavelengths and in Fig.\,\ref{CVF_Spectra} we show the background subtracted spectrum of the feature and there is no significant emission at shorter wavelengths. The close spatial correlation between the 16\,$\mu$m emission and the polarized light  (Fig.\,\ref{CVF_Image}) cannot be a coincidence and should carry information on the optical properties of the dust particles. 

One important question, not yet addressed, is the gas component in the disk-like feature. We failed to detect H$\alpha$, but this was expected in view of the low Lyman continua of B5\,V stars. Our search for H$_2$ fluorescence was not very deep and can certainly be improved. It may also be possible to detect the \ion{Ca}{ii} H and K lines in resonance scattering in a similar way to how it was detected in the disk around $\upbeta$\,Pic \citep{Olofsson2001} by means of long-slit, high resolution spectroscopy. An interesting aspect of our results in the present paper is the conclusion that -- be it a disk or a filament -- the feature should be located in the line of sight towards the star and the gas component should be seen in absorption. Our high resolution spectra show the expected interstellar absorption lines and diffuse interstellar bands (DIBs). There is only one velocity component (at $v_\mathrm{helio}$ = 14 km/s) and as this is the same as the radial velocity for the cloud, the interstellar absorption should arise locally in the foreground of the binary star. This means that any gas in the disk-like feature contributing to the observed absorption lines must share the radial velocity of the cloud.

\section{Conclusions}\label{sec:con}

We have shown that the disk-like feature exhibits a large contrast in polarized light and the high polarization degree, achieved after subtraction of the background, shows that the dust giving rise to the feature must be located in the vicinity of the star. A line drawn along the ridge of the feature passes very close to the direction of the star, which in principle can be a coincidence but taken together, these two findings indicate a physical connection between the feature and the central binary star. However, if the feature is due to a circumstellar ring it must have a non-uniform density distribution with a sector NW of the stars having a void of dust. If, after all, the feature is due to a narrow filament we show that the polarization results can be explained assuming a certain orientation of the filament.

The bright SE part of the disk-like feature as seen in polarized light perfectly coincides with the ISO/CVF image at 16\,$\mu$m.  The feature is, however not seen above the bright background in any of the PAH peaks at shorter wavelengths. This fact indicates that any PAH molecules that may have been hosted in the dust feature have disappeared, probably blown away by the radiation pressure. This means that the feature must have been close to the star long enough for such a process to be efficient. We have not estimated the required time scale, but it may add an additional constraint favoring the ring hypothesis.  

We show that at least one of the visual binary stars is a spectroscopic binary and that each of the SB components have $v\sin(i)$ values typical for B stars. The velocity separation at the time of our high resolution spectroscopy was large, 307 km/s, indicating that the SB should have a short period (days) and our intermediate spectroscopy during five consecutive nights support this interpretation. The system should be monitored spectroscopically (slit spectroscopy, including the two binary stars) for assessing the orbital parameters.

\begin{acknowledgements}
      B.\ Montesinos has kindly supported this work by making archive spectra available for us.
      We also thank A.\ Brandeker and R.\ Liseau for fruitful discussions on this and related topics.
      
      The PolCor instrument was funded by a grant from the Knut and Alice Wallenberg Foundation.
\end{acknowledgements}

\bibliographystyle{bibtex/aa}
\bibliography{myref} 


\begin{appendix}

\end{appendix}

\listofobjects

\end{document}